# Asymmetric response to PMI announcements in China's stock returns

Yingli Wang[1], Xiaoguang Yang[1]

1. Chinese Academy of Sciences, Academy of Mathematics and Systems Science, Beijing

E-mail: yinglwang5-c@my.cityu.edu.hk; xgyang@iss.ac.cn

**Abstract**

Considered an important macro-economic indicator, the Purchasing Managers' Index (PMI) on Manufacturing generally assumes that PMI announcements will produce an impact on stock markets. International experience suggests that stock markets react to negative PMI news. In this research, we empirically investigate the stock market reaction towards PMI in China. The asymmetric effects of PMI announcements on stock market are observed: no market reaction is generated towards negative PMI announcements, while a positive reaction is generally generated for positive PMI news. We further find that the positive reaction towards the positive PMI news occurs 1 day before the announcement and lasts for nearly 3 days, and positive reaction is observed in the context of expanding economic conditions. By contrast, the negative reaction towards negative PMI news is prevalent during downward economic conditions for stocks with low market value, low institutional shareholding ratios or high price earnings. Our study implies that China's stock market favors risk to a certain extent given the vast number of individual investors in the country, and there may exist information leakage in the market.

## 1. Introduction

In financial economics, the efficient market hypothesis states that asset prices fully reflect all available information, and prices can only move in response to news [1]. As such, we take the view that news on macroeconomic factors such as GDP, CPI, PPI and employment rate will influence financial markets, considering that the news announcements carry information about the aggregate investment opportunities targeted for the economy [2-7].

If positive (negative) macroeconomic news is announced, then investors may rebalance their portfolios by buying (selling) stocks, and this phenomenon will result in the rise (fall) of the stock market. If the positive and negative economic news have equal importance, then symmetric effects in the stock prices will be observed. However, if positive–negative asymmetry exists, then the stock market reaction towards the positive/negative economic news shocks will be unequal (i.e. the direction of asymmetry will depend on the relative 'importance' of the positive/negative news). The positivity (negativity) effect implies a special type of asymmetry. In particular, the positive (negative) reaction towards positive (negative) news will lead to a negligible reaction towards negative (positive) news.

There is ample evidence to conclude that a relationship between economic news and stock prices exists and that asymmetric stock market reactions accord with various types of news announcements [8-10]. A version of this finding is the 'negativity effect' (from the psychology literature) which can be described as follows: upon announcement of bad (good) sentiment news, the equity market experiences a significant negative news (no news) day effect [10]. The impact of economic news on equity appears to be nonlinear and asymmetric [11]. Stocks with bad public news display a negative drift for nearly 12 months, but less drift is found for stocks with good news [12]. In particular, negative news from the US market can cause a larger decline in national stock return compared with an equal magnitude of good news [13]. The stock market responds differently to positive and negative target rate surprises [14]. Firms generally react quickly to negative macroeconomic news, but some of them with few stocks can adjust to positive economic news with delay [15]. Subsequently, the volume and volatility effects of 21 macroeconomic news announcements on the S&P100 stock index options are considered, and the result shows negative news is associated with higher volume and volatility than positive news [16]. The negative aviation disaster is reported to cause more than $60 billion worth of stock market losses [17]. International sporting results are chosen as a proxy for investor sentiment, finding that when a country loses a game unexpectedly in an international sporting event, the losing country's stock market will fall significantly the following day, but the results did not show a corresponding increase in the winning country's stock market [18].

However, important macroeconomic news in most cases does not generate any effect. For example, various news on industrial production, unemployment and real GNP appear to have no significant impact on stock prices [19]. Similarly, markets largely disregard statistical announcements on industrial production and unemployment, even after prior expectations are taken into the account [20, 21]. Additionally, no significant market reaction towards GDP and unemployment news is detected [22].

In general, the different stages of the business cycle or economic conditions may change the context by which macroeconomic signals are interpreted. Subsequently, in the following studies, the market impact of news may become more apparent once the current economic conditions are considered. The effect of monetary news depends on market conditions (bull versus bear market), and monetary policy actions during bear market periods have a larger effect on stocks [23]. News on the rising unemployment has a negativity effect on stock prices during economic contractions because the announcements can indicate the lowering of corporate earnings and dividends; however, the positivity effect can be observed during expansions because good announcements can signal a greater likelihood of lower interest rates [24, 25]. In addition, after controlling for both market expectations and business cycle stages, the market may react significantly to GDP and unemployment announcements [26].

In this research, we choose the monthly manager index on manufacturing (i.e. PMI) announcements as the economic news. PMI announcements are typical representations of economic index news in China, and they serve as a timely indicator of economic forecasting and business analyses for the government, financial institutions, and companies. Furthermore, many scholars have shown the important effect of PMI news on the securities market. Findings have shown that PMI announcements impact the new stock markets of the European Union (e.g. Czech Republic, Hungary and Poland), that is, a worse-than-expected outcome provokes a negativity effect on the stock returns and vice versa [27]. Additionally, PMI changes have a greater impact on the stocks of smaller market capitalization firms and industries such as precious metals, computer technology, textiles, and automobiles [28]. Improvements to the PMI also affect the regime-switching probabilities during bull and bear stock market periods [29]. PMI news also significantly affects stock market volatility of the Group of Seven countries [30]. The effects of PMI announcements on the commodity futures indices, S&P 500 index, and government bond indices, including those in the US, has been established [31]. PMI news also affects the forward rates of several countries [32].

Based on the above literature, economic news is an important influencing factor of stock prices. Most of the above studies have focused on mature and well-established markets in the US and Europe. However, little attention has been devoted to the stock market of China. Notably, the Chinese stock market is the only national economic-driven state-owned-dominated stock market in the world and thus has many unique characteristics, such as high ratio of individual investors and strong government regulation. The distinct characteristics suggest that the conclusions offered in the literature on the mature US or European markets may not be applicable to the Chinese market. Consequently, the study on the stock market of China is particularly important and urgent.

The purpose of this research is to investigate the reaction of the managers' index (MI) stock returns towards PMI news and offer a number of suggestions to investors. The four goals are as follows. Firstly, we assess the relationship between surprise PMI and MI stock market behaviour. Secondly, we determine the lengths of the effects of the positivity or negativity. Thirdly, we establish whether the positivity or negativity effect of PMI news depends on economic conditions. Finally, we assess whether the result can be explained by different stock characteristics.

The remainder of the paper is structured as follows. Section 2 provides an overview on the data collection process. Section 3 presents the descriptive statistics and the results of the empirical regressions. Section 4 focuses on positivity-prone or negativity-prone portfolios. Section 5 discusses the robustness of the findings. Section 6 concludes the paper.

## 2. Data

The PMI is a set of economic indicators derived from the monthly surveys of companies conducted by the National Bureau of Statistics (NBS) of China and the China Goods Circulation and Purchasing League. A total of 727 enterprises participate in the PMI business survey. The questionnaire survey is conducted monthly with the purchasing managers of the participating enterprises. The PMI indicator system includes the following 11 indices: new order form, production, employment, delivery of suppliers, stock-in-trade, new exporting order form, purchasing, finished product stock, purchasing price, imports, and overstock order form. The PMI is a good indicator of the economic health of the manufacturing sector. The PMI system is established to provide information about the current business conditions to company decision makers, analysts and purchasing managers.

The PMI survey is significant to macro-economic and industrial economic adjustments, control and forecast, and thus can guide purchasing, production and operation at the enterprise level. If the PMI contains such valuable information and the market is efficient, then the PMI information is expected to be incorporated into the manufacturing stock prices by the end of the announcement day. The monthly PMI announcements in the present study are from January 2005 to March 2018 (end of sampling period). The PMIs are announced to the market at 9 a.m. on the first day of each month, and investors have public access to the data during the day.

On the basis of previous research, if the change in the news relative to the previous month is greater (lesser) than 0, then the news of the month selected is called positive (negative) news; otherwise, the change is called no news [10, 33]. However, the PMI in China is also unique relative to those of other countries. If the PMI is above 50, then the manufacturing economy has the status of general expansion. If the PMI is below 50, then the manufacturing economy has the status of generally recession. In this sense, the PMI operates as a timely indicator of economic forecasting and business analysis for the government, financial institutions, and companies. By taking into consideration the literature categorisation and the Chinese reality, we give new definitions of positive news, negative news and no news in this research as follows. If the selected PMI news is higher than that of the previous month and more than 50, then it is a positive news. If the PMI news is lower than that of the previous month and less than 50, then it is a negative news. Otherwise, the PMI news is called no news.

The Shenzhen MI is an industry index published by Shenzhen Stock Exchange since July 2, 2001, and it can represent comprehensive stock price trends of the manufacturing industry. Considering that PMI and MI are both associated with the manufacturing industry, we adopt these two indices to examine the relationship of economic news and stock price movement from the perspective of manufacturing.

We obtain the daily values from the Wind dataset for a range of items, including MI as proxy of the market returns of the manufacturing industry and daily returns of Shanghai Stock Exchange (SSE) Composite Index for MI control. The manufacturing PMI data are confirmed by NBS-China. The period of January 2005 to March 2018 comprise approximately 2,700 stock trading days over a period of 159 months. Thus, the corresponding 159 monthly manufacturing PMIs and 2,700 MI, SSE Composite Index daily returns are gathered for the sample period.

## 3. Empirical analysis
### 3.1 Descriptive statistics

In this subsection, we present basic statistics and notations which are necessary for the understanding of subsequent results.

Firstly, we derive the specific statistics on the daily returns of MI for the announcement days. Table 1 presents the details on the positive, negative and no-change PMIs. The results indicate an asymmetric market effect of the PMI news on the MI stock returns. If the value of the PMI of a selected month relative to the previous month is increased (decreased) and the value is greater (lesser) than 50, then we call it positive (negative) PMI; otherwise, it is a no-change PMI. In Figure 1, the period of January 2005–May 2018 covers 159 months. During these 159 announcement days, 67 days have positive PMIs, 11 have negative PMIs, and 81 days have no-change PMIs.

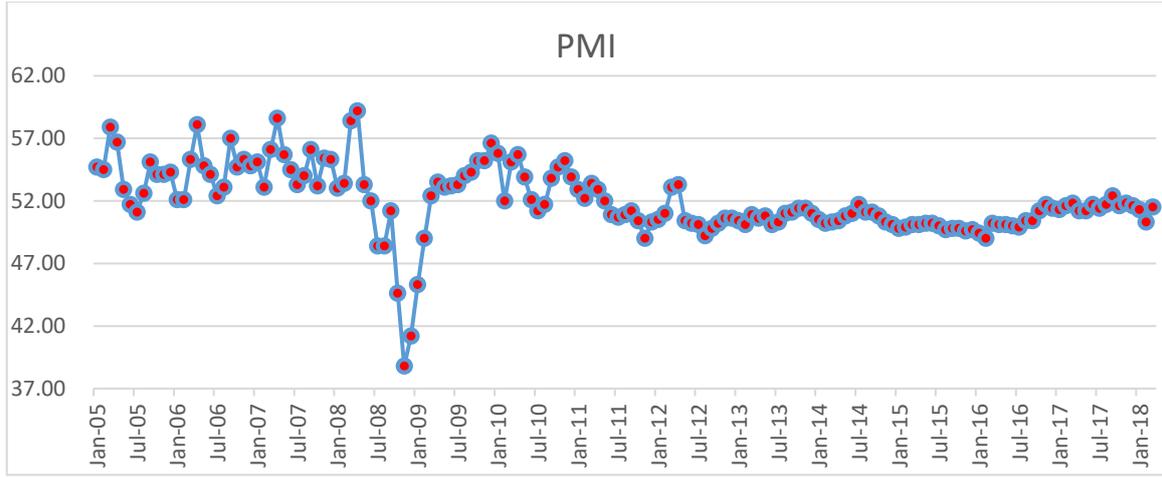

**Figure 1 Monthly PMIs**

Table 1 presents the average MI returns of the different PMI announcement days, in which a series of t-tests is conducted to compare the mean values of the three groups. The average daily MI returns for the announcement days with positive, negative and no-change PMIs are 0.863%, −0.073% and 0.08%, respectively. Here, MI return for day t is calculated by $\log(\frac{MI_t}{MI_{t-1}})$, and $MI_t$ is the closing price of MI on day t. On the basis of the t-test results, the average returns for the announcement days with positive versus negative PMIs reject the null hypothesis of zero difference (P-value = 0.0068); for those between positive and no-change PMIs, the relationship is significant (P-value = 0.0014); and for those between negative and no-change PMIs, the relationship is not significant (P-value= 0.9021). These findings support the argument on the positivity effect of PMIs on MI stock returns.

**Table 1 Average daily returns of MI for announcement days relative to PMI types**

| Variable | Positive PMI | Negative PMI | No-change PMI |
|---|---|---|---|
| Average daily return of MI | 0.863% | −0.073% | 0.08% |
| P-value on difference: MI returns on announcement day of PMI | positive vs. no-change PMI: 0.0014 | negative vs. no-change PMI: 0.9021 | positive vs. negative PMI: 0.0068 |

### 3.2 Baseline regression model

To estimate the impact of the type and change of PMI on MI returns, we estimate the following model:

$$R_t = \beta_0 + \beta_1 Pos_t + \beta_2 Neg_t + \beta_3 R_{m,t} + \varepsilon_t, \quad (1)$$

where $R_t = \log(\frac{MI_t}{MI_{t-1}})$ is the continuous daily return on MI, while $MI_t$ is the closing price of MI for day t. $Pos_t$ and $Neg_t$ are dichotomous variables. In particular, $Pos_t$ takes a value of the unity for the day that is associated with the positive PMI announcement, while $Neg_t$ is the direct counterpart for the negative PMI announcement; otherwise, the value is 0. $R_{m,t} = \log(\frac{SSE_t}{SSE_{t-1}})$ represents returns on Chinese stock market, where $SSE_t$ is the closing price of SSE Composite Index on day t.

Table 2 presents the calculations of Equation (1). In Panel A, the estimated coefficient on the positive change in $Pos_t$ is positive and significant at the 5% level. By contrast, the estimated coefficient on the negative change in $Neg_t$ is not significant. These results support the statistics in Table 1 which suggest the existence of the positivity effect.

To further confirm the positivity effect on the MI returns, we extend the samples to the 1,195 stocks returns of the MI to produce a new set of 89,303 samples. The regression results of extended samples are presented in Panel B of Table 2, where $\beta_1$ is significantly positive, while $\beta_2$ is insignificantly negative, such confirming the positivity effect.

**Table 2 Asymmetric effects of PMI news on stock returns**

| Panel A: PMI news modelled by the 159 MI stock returns on announcement days | | | | | | |
|---|---|---|---|---|---|---|
| | $\beta_0$ | $\beta_1$ | $\beta_2$ | $\beta_3$ | F-stat | Adj. $R^2$ |
| estimate | 0.000 | 0.008 | 0.003 | 0.130 | 1.540 | 0.012 |
| t-Stat | 0.06 | 1.91 | 0.37 | 1.13 | | |
| p-Value | 0.955 | 0.058 | 0.711 | 0.261 | 0.207 | |
| Panel B: PMI news modelled by 89,303 specific MI stock returns on announcement days | | | | | | |
| | $\beta_0$ | $\beta_1$ | $\beta_2$ | $\beta_3$ | F-stat | Adj. $R^2$ |
| estimate | −0.007 | 0.005 | −0.001 | 0.007 | 274.03 | 0.041 |
| t-Stat | −18.12 | 18.41 | −1.280 | 0.75 | | |
| p-Value | <0.000 | <0.000 | 0.200 | 0.451 | <0.000 | |

### 3.3 Event study

An event study is conducted to further explore the dynamics of the positivity effect. Table 3 presents the calculated daily average abnormal returns (AARs) when time is set to −3, −2, −1, 0, 1, 2 and 3. For the detailed calculation of average abnormal returns (AARs), please refer to ****. Here, 0 is designated as the announcement day of PMI, while −1 and 1 are the days before and after the announcement day, respectively.

As depicted by Panel B of Table 3, the effect of announcement days on negative PMI events across all event windows is not significant. However, concerning positive PMI announcements, the effect on the event windows is significantly positive. As depicted by Panel A of Table 3, in terms of the AARs of MI with positive PMI news, four notable time-events (−1, 0, 1 and 2) can explain the significant outcomes at the <10% level (0.67%, 0.86%, 0.61% and 0.53% returns, respectively).

Specifically, the information on the positive PMI announcement has been incorporated into the stock market prices the day before and during the announcement day, but no similar effect is observed for negative PMI news. Furthermore, the MI market has recovered from the shock post-announcement of the positive PMI news. The AARs of MI are insignificant on the third day of after the announcement. This finding is consistent with the market having to rebalance portfolios and stocks within 3 days after the positive announcement of PMI.

**Table 3 AARs for different event days for positive and negative PMI announcements**

| Panel A: AARs for positive PMI announcements ($H_0 = 0$) | | | | | | | |
|---|---|---|---|---|---|---|---|
| event day | −3 | −2 | −1 | 0 | 1 | 2 | 3 |
| AARs of MI | −0.004 | −0.001 | 0.007 | 0.009 | 0.006 | 0.005 | 0.003 |
| t-Stat | −1.85 | −0.17 | 0.64 | 2.84 | 2.01 | 1.99 | 1.55 |
| p-Value | 0.070 | 0.864 | 0.022 | 0.006 | 0.050 | 0.051 | 0.128 |
| Panel B: AARs for negative PMI announcements ($H_0 = 0$) | | | | | | | |
| event day | −3 | −2 | −1 | 0 | 1 | 2 | 3 |
| AARs of MI | −0.003 | −0.005 | −0.009 | −0.001 | −0.004 | 0.004 | 0.004 |
| t-Stat | −0.58 | −0.44 | −1.13 | −0.09 | −0.51 | 0.50 | 1.07 |
| p-Value | 0.579 | 0.67 | 0.291 | 0.929 | 0.621 | 0.633 | 0.316 |

### 3.4 Interaction between PMI news and economic conditions

The preceding studies assumed that economic conditions generally do not affect or interact with PMI. In fact, many investors regard economic conditions as contagious, so such that several investors may shift their actions in accordance with those interpreted conditions. Thus, the topic of effect of PMI under different economic conditions is worth the investigation.

In this part of the study, we test whether positive or negative news can lead to stronger market response by controlling for economic conditions. If yes, is there asymmetry in this relationship? The aim of the regression is to answer this particular question.

Here, we use the Chinese Economic Sentiment Index as the indicator of economic situation. The Economic Sentiment Index is obtained from the Global Economic Monitoring and Policy Simulation Platform published by the Chinese Academy of Sciences. The overall aim of the Economic Sentiment Index is to provide a method to assess the strengths and weakness of the Chinese economy. Thus, the higher the Economic Sentiment Index is, the better the economy becomes, and vice versa. Figure 1 shows the Economic Sentiment Index of China from January 2005 to March 2018. We consider the months with up-sloping curves to be up-economic times and the months with down-sloping curves to be down-economic times. That is to say, if the Economic Sentiment Index this month is higher (lower) than former one, then this month is up (dowm)-economic time.

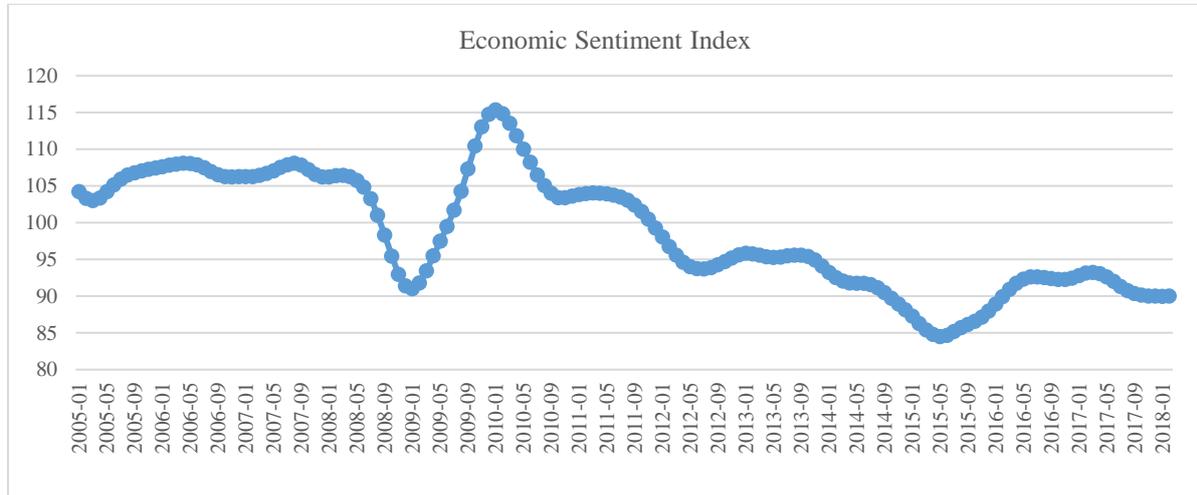

**Figure 2 Economic Sentiment Index of China**

To explore the question as to how PMI and economic conditions interact with one another, we adopt a simple but intuitive approach. Let the $2 \times 2$ dimensional representation be the four potential cases of (1) positive PMI change/up-economic condition, (2) negative PMI change/up-economic condition, (3) positive PMI change/down-economic condition and (4) negative PMI change/down-economic condition. The model can then be expressed as

$$R_t = \beta_0 + \beta_{1u} Pos_t\ Up_t + \beta_{2u} Neg_t\ Up_t + \beta_{1d} Pos_t\ Down_t + \beta_{2d} Neg_t\ Down_t + \beta_3 R_{m,t-1} + \varepsilon_t, \quad (2)$$

where $Up_t$ ($Down_t$) is a dichotomous variable that takes a value of unity if the economic index shows a positive (negative) increase relative to the economic index of the previous month; otherwise, the value is 0. The other variables have been previously defined in this paper. The results for estimating this model are shown in Table 4.

**Table 4 Effect of positive and negative PMI news on MI returns in different economic conditions**

|          | $\beta_{1u}$ | $\beta_{2u}$ | $\beta_{1d}$ | $\beta_{2d}$ | $\beta_3$ | F-stat |
|----------|-------|--------|-------|-------|-------|--------|
| estimate | 0.011 | −0.005 | 0.006 | 0.011 | 0.145 | 1.28   |
| t-Stat   | 2.05  | −0.42  | 1.05  | 0.99  | 1.24  |        |
| p-Value  | 0.043 | 0.672  | 0.294 | 0.324 | 0.218 | 0.075  |

Table 4 confirms the assumption on the positive change in PMI news in up-economic conditions, that is, the estimated coefficient of the positive/up movement that is statistically significant. This result supports the positivity effect, albeit in conditional form. Notably, the 'conflicting interaction' terms (i.e. positive/down, negative/up, negative/down) did not show statistical significance.

From the above analysis, financial investors seem to be only affected by positive news in up-economic conditions. Even if PMI decreases, investors tend to neglect the negative news and would not drastically change their behaviour. Moreover, the investors are insensitive to any news (positive or negative) in down-economic conditions. In summary, the positive PMI effect can be observed in up-economic times only (i.e. after taking economic conditions into the account).

## 4. Positivity-prone or negativity-prone portfolios

As discussed in the previous section, the asymmetric effect of PMI news on MI returns is apparent, and investors are likely to gain profits from buying stocks in the MI market on the day the positive announcement is made. However, there are more than 1,000 stocks with different characteristics in the MI market, and determining which one to buy is still a question. As such, our next analysis is focused on the investors' decision on which kinds of stocks to buy (sell) upon receiving the positive (negative) PMI announcement.

We argue that if investors rebalance their portfolios upon the announcement of positive (negative) PMI news, they are likely to be selective on which stocks to buy (sell). We also argue that the asymmetric effect is more likely observable across firms with different sizes, institutional shareholding ratios, or growth potentials. Thus, in this part of the study, we create alternative sets of test portfolios to explore how strongly the asymmetric effect will be on portfolios formed on the basis of (1) size, (2) institutional shareholding ratio and (3) price-earning (P/E).

The small (large) firm portfolio comprises 20% of sample stocks with the smallest (largest) sizes on the basis of the annual rankings on market equity capitalisation. The low (high) institutional shareholding ratio portfolio comprises 20% of sample stocks with the lowest (highest) institutional shareholding ratios based on the annual ranking of institutional shareholding ratio. The low (high) P/E portfolio comprises 20% of sample stocks with the lowest (highest) growths based on the annual ranking of firm P/E ratios. Previous studies have shown that economic conditions affect or interact with PMI news. Thus, we take the interaction of economic condition and PMI news into consideration.

Table 5 presents the estimated outcome of Eq. (2) for the extreme portfolios formed on the basis of the above three firm characteristics. Panel A depicts our results for the small and big firm size portfolios (market value). Interestingly, the estimated coefficient for the positive PMI dummy in the up-economic condition is significant, whereas the negativity effect is only significant in the down-economic condition for the small-firm portfolio. Panel B of Table 5 shows the analytical results of the portfolios with low and high institutional shareholding ratios. While both portfolios in the up-economic conditions exhibit the positivity effect, the negativity effect only appears in the low institutional shareholding ratio portfolio in down-economic conditions. Panel C of Table 5 presents the corresponding results for the portfolios in terms of P/E. The portfolios with high P/E have a positive and significant estimated coefficient for the positive PMI dummy in the up-economy condition with a p-value of <0001. However, both low and high institutional P/E portfolios do not reveal any statistical significance in relation to negative PMI news.

As shown in Table 5, firms with different types or characteristics react differently towards the positivity or negativity effect. Firms with low market values or low institutional shareholding ratios show both positivity and negativity effects in up-economic and down-economic conditions, respectively. Firms with high market values or high institutional shareholding ratios in up-economic conditions also show a positivity effect. Firms with low P/E are affected neither by the positivity nor negativity, and only the firms with high P/E in up-economic conditions experience the positivity effect. As such, when faced with positive PMI news in up-economic conditions, investors should avoid buying stocks with low P/E. By contrast, when confronted with negative PMI news in down-economic conditions, investors need to sell stocks of low market value or with low institutional shareholding ratio.

Table 5 Effects of positive and negative PMI news on stocks with different characteristics

| Panel A: Market value (20%) | | | | | | |
|---|---|---|---|---|---|---|
| **Low** | $\beta_{1u}$ | $\beta_{2u}$ | $\beta_{1d}$ | $\beta_{2d}$ | $\beta_3$ | F-stat |
| estimate | 0.006 | 0.001 | 0.005 | −0.005 | −0.085 | 89.08 |
| t-Stat | −3.46 | −0.42 | 1.35 | −4.78 | −4.45 | |
| p-Value | 0.001 | 0.201 | 0.171 | <.000 | <.000 | 0.075 |
| **High** | $\beta_{1u}$ | $\beta_{2u}$ | $\beta_{1d}$ | $\beta_{2d}$ | $\beta_3$ | F-stat |
| estimate | 0.009 | 0.010 | 0.014 | 0.001 | 0.099 | 51.46 |
| t-Stat | 13.55 | −0.32 | 0.25 | 0.48 | 5.02 | |
| p-Value | <0.000 | 0.167 | 0.311 | 0.631 | <0.001 | <0.000 |
| **Panel B: Institutional shareholding ratio (20%)** | | | | | | |
| **Low** | $\beta_{1u}$ | $\beta_{2u}$ | $\beta_{1d}$ | $\beta_{2d}$ | $\beta_3$ | F-stat |
| estimate | 0.002 | 0.001 | −0.000 | −0.003 | 0.077 | 50.18 |
| t-Stat | 2.38 | 3.45 | 1.02 | 2.34 | 3.66 | |
| p-Value | 0.018 | 0.45 | 0.245 | 0.020 | <0.000 | <0.000 |
| **High** | $\beta_{1u}$ | $\beta_{2u}$ | $\beta_{1d}$ | $\beta_{2d}$ | $\beta_3$ | F-stat |
| estimate | 0.009 | −0.002 | 0.006 | −0.001 | −0.032 | 72.39 |
| t-Stat | 13.98 | −0.32 | 0.05 | −1.09 | −1.61 | |
| p-Value | <0.000 | 0.654 | 0.344 | 0.277 | 0.107 | <.000 |
| **Panel C: P/E (20%)** | | | | | | |
| **Low** | $\beta_{1u}$ | $\beta_{2u}$ | $\beta_{1d}$ | $\beta_{2d}$ | $\beta_3$ | F-stat |
| estimate | 0.001 | −0.003 | 0.004 | −0.001 | −0.019 | 64.42 |
| t-Stat | 1.51 | −0.65 | 1.23 | −1.11 | −0.99 | |
| p-Value | 0.130 | 0.632 | 0.305 | 0.267 | 0.324 | <0.000 |
| **High** | $\beta_{1u}$ | $\beta_{2u}$ | $\beta_{1d}$ | $\beta_{2d}$ | $\beta_3$ | F-stat |
| estimate | 0.007 | −0.001 | 0.007 | −0.000 | 0.054 | 47.75 |
| t-Stat | 11.47 | −0.12 | 1.35 | −0.14 | 2.67 | |
| p-Value | <0.000 | 0.782 | 0.452 | 0.890 | 0.008 | <0.000 |

## 5. Robustness check

In this part of the study, we investigate the robustness of the positive PMI effect by considering a range of issues. One of the concerns is the definition of positive and negative PMIs, while the other issue is related to the selection of economic news. Our findings on the positivity effect of economic news on stock returns suggest that our research method is robust.

The first issue is the definition of positive and negative PMIs. In the earlier sections of this paper, we defined $Pos_t$ ($Neg_t$) as dichotomous variables that take the value of unity for the day of an announcement when the adjusted PMI relative to the previous month is increased (decreased), and that the value of the PMI is larger (smaller) than 50 (otherwise, 0). Positivity may have different definitions to investors. Thus, we take two other kinds of definitions of positive and negative PMIs to test the positivity effect. The first definition is used only for the PMI of the previous month. If $PMI_t - PMI_{t-1}$ is larger (smaller) than 0, then $Pos_t$ ($Neg_t$) takes the value of unity; otherwise, the value is 0. Panel A in Table 6 presents the estimated results of Eq. (1) by using the new definition of positive PMI. Meanwhile, the second definition is used to compare the average PMI of the three previous months. If $PMI_t - (PMI_{t-1} + PMI_{t-2} + PMI_{t-3})/3$ is larger (smaller) than 0.5, then $Pos_t$ ($Neg_t$) takes the value of unity; otherwise, the value is 0. The average increase in value of the PMI is 1.384, while the average decrease in value is −1.408. To eliminate collinearity, we take [−0.5, 0.5] for the value of the no-change PMI. The estimation by using this new definition of positive PMI is tested with Eq. (1). The results are shown in Panel B of Table 6. Then, a follow-up analysis is conducted. The results show that our previous findings on the positivity effect are robust to the change in the first definition. For the second definition of positive PMI, we take [−0.25, 0.25] as the value of no-change PMI. The results show that only the positivity effect is significant.

**Table 6 Positivity effect with different definitions of positive and negative PMIs**

| **Panel A:** $Pos_t=1$ for $PMI_t - PMI_{t-1} > 0$ | | | | | | |
|---|---|---|---|---|---|---|
| | $\beta_0$ | $\beta_1$ | $\beta_2$ | $\beta_3$ | F-stat | Adj. $R^2$ |
| estimate | −0.011 | 0.019 | 0.013 | 0.163 | 2.55 | 0.0344 |
| t-Stat | −1.53 | 2.46 | 1.67 | 1.45 | | |
| p-Value | 0.127 | 0.015 | 0.17 | 0.151 | 0.0583 | |
| **Panel B:** $Pos_t=1$ for $PMI_t - (PMI_{t-1} + PMI_{t-2} + PMI_{t-3})/3 > 0$ | | | | | | |
| | $\beta_0$ | $\beta_1$ | $\beta_2$ | $\beta_3$ | F-stat | Adj. $R^2$ |
| estimate | −0.000 | 0.009 | 0.003 | 0.108 | 1.54 | 0.012 |
| t-Stat | −0.03 | 1.88 | 0.61 | 0.97 | | |
| p-Value | 0.973 | 0.062 | 0.541 | 0.333 | 0.208 | |

The second issue is related to the alternative modelling of economic news. As previously shared, the aim of this research is to test if the positive macroeconomic news has a stronger effect on the stock market compared with the negative news. Thus, to test the positivity effect, we use another economic parameter that can depict the growth rate of industry value-adding instead of the PMI news. The Industrial Added Value index refers to the final results of industrial enterprise production, and its monetary terms corresponding to the study period are considered. The results of the estimation are shown in Table 7.

**Table 7 Positivity effects of the value-added feature of industry news**

| | $\beta_0$ | $\beta_1$ | $\beta_2$ | $\beta_3$ | F-stat | Adj.$R^2$ |
|---|---|---|---|---|---|---|
| estimate | −0.005 | 0.007 | 0.006 | 0.021 | 1.40 | 0.006 |
| t-Stat | −1.21 | 1.65 | 1.46 | 0.29 | | |
| p-Value | 0.227 | 0.081 | 0.147 | 0.335 | 0.049 | |

Table 7 shows the estimated outcome of Eq. (1). In general, our previous finding on the robustness of the positivity effect is confirmed. In particular, as shown in Table 7, the estimated coefficient of the positive change for the value-added feature of industry news remains positive and significant at the level of 8.1%. The corresponding estimated coefficient of the negative change is insignificant.

## 6. Conclusion

This research analyses the positivity effect of PMI announcements on MI market returns by using the index issued by NBS-China. Firstly, when the announced PMI index is higher than that of the previous month and larger than 50, the MI returns exhibit a significant positive announcement-day effect. However, when the announced PMI index is lower than that of the previous month and lesser than 50, no significant effect on MI returns is observed. Secondly, the MI market recovers from 'positive PMI news' shock within 3 days. In addition, economic conditions are determinants of PMI. In particular, PMI news that is positive and accords with up-economic conditions have significant positivity effects on MI returns. However, the 'conflicting interaction' terms (positive/down, negative/up, negative/down) do not show any statistical significance. Furthermore, firms with different characteristics have varying positivity-prone or negativity-prone characteristics. The findings suggest that our research method is robust. The robustness is further confirmed by the two other types of definitions on positive (negative) PMI news and the alternative modelling of economic index news.

Our research, as well as the literature body to which it belongs, shows that news with economic relevance are important aspects of stock price movements. An important contribution of our work is the documentation of the asymmetric effect that combines with the short-term effect of PMI news. In addition to economic announcements, the uneven manner in which markets generally react to information provides interesting avenues for future research.